# Quantum Detectability in Invisibility Cloaks

Mohammad Mehdi Sadeghi
[a] Department of Physics, Jahrom University, Jahrom, Iran
* *Corresponding author e-mail: sadeghi@jahromu.ac.ir*

**Abstract** :Classical invisibility cloaks are designed to suppress selected scattering signatures and thereby make an object appear absent to external electromagnetic probes. However, the suppression of a classical scattering observable does not, by itself, establish that all information about the concealed object has been removed from the detected quantum state of light. Here we formulate the detectability of classically cloaked objects as a quantum-state distinguishability problem. Treating a linear passive cloak as an effective Gaussian quantum channel acting on the accessible detected modes, we show that local quantum undetectability requires the detected first and second moments to be independent of the hidden-object parameter. In this framework, quantum Fisher information provides an operational criterion for whether the concealed parameter remains estimable from the detected output state. We derive displacement- and covariance-level detectability conditions and show that a nonzero parameter imprint surviving in the detected Gaussian state leads to a nonzero accessible quantum Fisher information. To connect the criterion with a physical cloaking model, we analyze a regularized cylindrical transformation-optical cloak in the Born limit and compare the scaling of the classical scattering response with the derivative-based quantum sensitivity. The analysis shows that reducing a scattering amplitude is not equivalent to eliminating local quantum-state sensitivity. Loss, environmental noise, and finite numerical aperture degrade the accessible information, but quantum undetectability is reached only when the parameter imprint is removed from the detected state or projected entirely outside the accessible subspace. These results provide a Gaussian-channel framework for assessing when classical cloaking does, and does not, imply quantum-state undetectability.

## 1. Introduction

Invisibility cloaking is commonly associated with the suppression of electromagnetic scattering from an object or region of space. Transformation optics provides a systematic route for designing anisotropic and inhomogeneous media that guide electromagnetic fields around a concealed domain, thereby reducing the external scattering signature of the hidden object [1–5]. This idea has led to a broad range of cloaking and wave-control schemes, including singular and regularized transformation-optical cloaks, carpet cloaks, plasmonic cloaks, optical illusion devices, null-media transformations, and scattering-cancellation approaches [6–11].

Most classical cloaking criteria are formulated in terms of field amplitudes, far-field patterns, scattering cross-sections, or image contrast [7,9,12–17]. These quantities are physically meaningful and experimentally accessible, but they test only selected observables of the outgoing field. They do not, by themselves, establish whether all information about the concealed object has been removed from the detected optical state. A cloak may strongly suppress a chosen scattering signature while leaving a residual parameter imprint in phase, quadrature displacement, covariance, loss-induced noise, or angular modal structure.

Here we formulate this distinction as a quantum-state detectability problem. If the concealed object is described by a parameter $\theta$, the relevant question is whether the detected output state $\rho_{\text{det}}(\theta)$ depends on $\theta$. In the Gaussian setting considered in this work, the detected state is specified by its first moment $\boldsymbol{\mu}_{\text{det}}(\theta)$ and covariance matrix $V_{\text{det}}(\theta)$. Local quantum undetectability therefore requires

$$\partial_\theta \boldsymbol{\mu}_{\text{det}}(\theta) = 0, \qquad \partial_\theta V_{\text{det}}(\theta) = 0$$

If either derivative is nonzero, the detected Gaussian state retains local information about the hidden object. Quantum Fisher information then provides the operational measure of how precisely this residual imprint can be estimated from the detected output state [24–31].

We model a linear passive cloak as an effective Gaussian quantum channel acting on the accessible detected modes. Absorption, finite collection, inaccessible radiation modes, and material reservoirs enter through the reduced input–output channel and its added noise [18–24]. This framework separates three issues that are often conflated: suppression of a classical scattering observable, removal of the hidden-object imprint from the detected quantum state, and degradation of the accessible information by loss, noise, and finite numerical aperture.

To connect the general criterion with a physical cloaking model, we analyze a regularized cylindrical transformation-optical cloak. Regularization replaces the singular compression of the ideal cloak by a finite effective virtual radius [12–14]. In a scalar cylindrical Born-limit benchmark, the exterior modal response can be related to the scattering coefficient of an effective virtual cylinder [15–17]. This allows a direct comparison between the classical scattering strength, which scales with $| s_m |^2$, and the derivative-based quantum sensitivity, which scales with $| \partial_\theta s_m |^2$. The comparison shows explicitly why reducing a scattering amplitude is not equivalent to eliminating local quantum-state sensitivity.

The paper is organized as follows. Section 2 distinguishes classical scattering invisibility from quantum-state undetectability. Section 3 formulates a linear passive cloak as an effective Gaussian quantum channel on the detected modes. Section 4 introduces the quantum Fisher information criterion for local Gaussian-state detectability. Section 5 develops the regularized cylindrical cloak benchmark. Section 6 analyzes the effects of loss, environmental noise, and finite numerical aperture. Section 7 discusses the scope and implications of the framework, and Section 8 summarizes the main conclusions. Appendix A gives the Gaussian-state quantum Fisher information used in the main text, while Appendix B provides the Born-limit derivation for the regularized cylindrical cloak.

## 2. Classical Cloaking versus Quantum-State Undetectability

Classical invisibility cloaking is usually defined through the suppression of an electromagnetic scattering signature [1–9]. If the concealed configuration is described by a parameter $\theta$, a typical classical condition may be expressed schematically as

$$\sigma_{\mathrm{sca}}(\theta) \to 0, \qquad (1)$$

where $\sigma_{\mathrm{sca}}$ denotes a selected scattering observable, such as a total scattering cross-section, an angular scattering distribution, or a far-field intensity contrast. Equation (1) captures the usual operational meaning of classical cloaking: the field measured in a chosen scattering channel becomes indistinguishable, or nearly indistinguishable, from that of a reference configuration.

However, Eq. (1) is not a statement about the full information content of the detected electromagnetic state. It constrains a selected classical observable, not the complete set of quantum-mechanical degrees of freedom available to a detector. In particular, an object-dependent imprint may remain in the optical phase, quadrature displacement, covariance matrix, noise correlations, polarization structure, frequency modes, or angular components outside the particular scattering observable used to define invisibility [24,32–41].

To formulate undetectability at the quantum level, let the detected output state be denoted by $\rho_{\mathrm{det}}(\theta)$. Two hidden-object configurations $\theta_0$ and $\theta_1$ are operationally indistinguishable on the detected subspace only if

$$\rho_{\mathrm{det}}(\theta_0) = \rho_{\mathrm{det}}(\theta_1) \qquad (2)$$

For a local estimation problem, in which the hidden parameter is varied infinitesimally around a reference value, quantum-state undetectability requires

$$\partial_\theta \rho_{\mathrm{det}}(\theta) = 0 \qquad (3)$$

Equation (3) is stronger than the classical condition in Eq. (1). A vanishing or strongly suppressed scattering cross-section does not generally imply that the entire detected quantum state is independent of $\theta$. Thus,

$$\sigma_{\mathrm{sca}}(\theta) \to 0 \;\; \not\Rightarrow \;\; \partial_\theta \rho_{\mathrm{det}}(\theta) = 0 \qquad (4)$$

Equation (4) is the central logical distinction used throughout this work. Classical cloaking suppresses a selected scattering signature, whereas quantum undetectability requires the absence of parameter dependence in the detected state itself.

For the Gaussian optical states considered below, the detected state is completely specified by its first moment $\mu_{\mathrm{det}}(\theta)$ and covariance matrix $V_{\mathrm{det}}(\theta)$ [24–31]. Therefore, the local condition in Eq. (3) becomes

$$\partial_\theta \mu_{\mathrm{det}}(\theta) = 0, \partial_\theta V_{\mathrm{det}}(\theta) = 0 \qquad (5)$$

If either derivative in Eq. (5) is nonzero, then the detected Gaussian state retains a local imprint of the hidden object. The role of quantum Fisher information, introduced in Section 4, is to quantify the ultimate sensitivity of the detected state to this imprint [32–37].

This distinction also clarifies the scope of the present work. We do not claim that an ideal mathematical cloak must always be detectable. If a cloak and detection geometry remove all dependence of $\rho_{\text{det}}(\theta)$ on the hidden parameter, then no measurement restricted to the detected subspace can recover $\theta$. The claim is instead that classical scattering suppression alone is insufficient as a certificate of quantum-state undetectability. To establish true quantum undetectability, one must verify the stronger state-independence condition in Eq. (3), or equivalently Eq. (5) for Gaussian detected states.

## 3. Effective Gaussian Channel of a Cloaked System

We now describe the detected output of a cloaked optical system as an effective quantum channel. The purpose is not to rederive the microscopic quantization of the electromagnetic field in dispersive and absorbing media, but to identify the operational input–output map through which information about the hidden parameter θ can reach the accessible detected modes. We assume throughout that the cloak is linear and passive. Linearity implies a linear transformation of field operators, while passivity excludes phase-sensitive amplification. Absorption, radiation into uncollected directions, finite numerical aperture, and material reservoirs are included through environmental modes, as required for a physically valid quantum input–output relation [18–23].

Let $\hat{\mathbf{a}}_{\text{in}}$ denote a vector of incoming annihilation operators in a chosen modal basis, and let $\hat{\mathbf{a}}_{\text{out}}(\theta)$ denote the corresponding outgoing modes that are accessible to the detector after interaction with the cloaked system. A general linear passive input–output relation may be written as

$$\hat{\mathbf{a}}_{\text{out}}(\theta) = T(\theta)\hat{\mathbf{a}}_{\text{in}} + L(\theta)\hat{\mathbf{b}}_{\text{env}}, \qquad (6)$$

where $T(\theta)$ is the coherent transmission matrix from the input modes to the accessible output modes, $L(\theta)$ describes coupling from unobserved environmental modes into the detected output, and $\hat{\mathbf{b}}_{\text{env}}$ is a vector of independent bosonic reservoir operators. The environmental term in Eq. (6) is not a phenomenological addition. It is required whenever absorption, mode elimination, or incomplete detection is present.

The output operators must satisfy the bosonic commutation relations

$$[\hat{a}_{\text{out},i}(\theta), \hat{a}^{\dagger}_{\text{out},j}(\theta)] = \delta_{ij} \qquad (7)$$

If the input and environmental modes are independent bosonic modes, Eq. (7) gives the passive-channel constraint

$$T(\theta)T^{\dagger}(\theta) + L(\theta)L^{\dagger}(\theta) = I \qquad (8)$$

In the ideal lossless and fully accessible limit, $L(\theta) = 0$, and Eq. (8) reduces to $T(\theta)T^{\dagger}(\theta) = I$. The transformation is then unitary on the full accessible modal space. Realistic detected cloaking channels are generally not unitary on the measured subspace, because finite collection, absorption, and inaccessible radiation modes remove part of the outgoing field from the detector [18–24].

Figure 1 illustrates this physical input–output structure. Incident modes interact with a cloak enclosing a hidden region characterized by θ. The outgoing modes may carry a parameter-dependent imprint, while loss, absorption, material reservoirs, and uncollected radiation are represented by unobserved environmental modes. The detector collects only a finite modal subspace, represented by the projection $P_{\mathrm{det}}$.

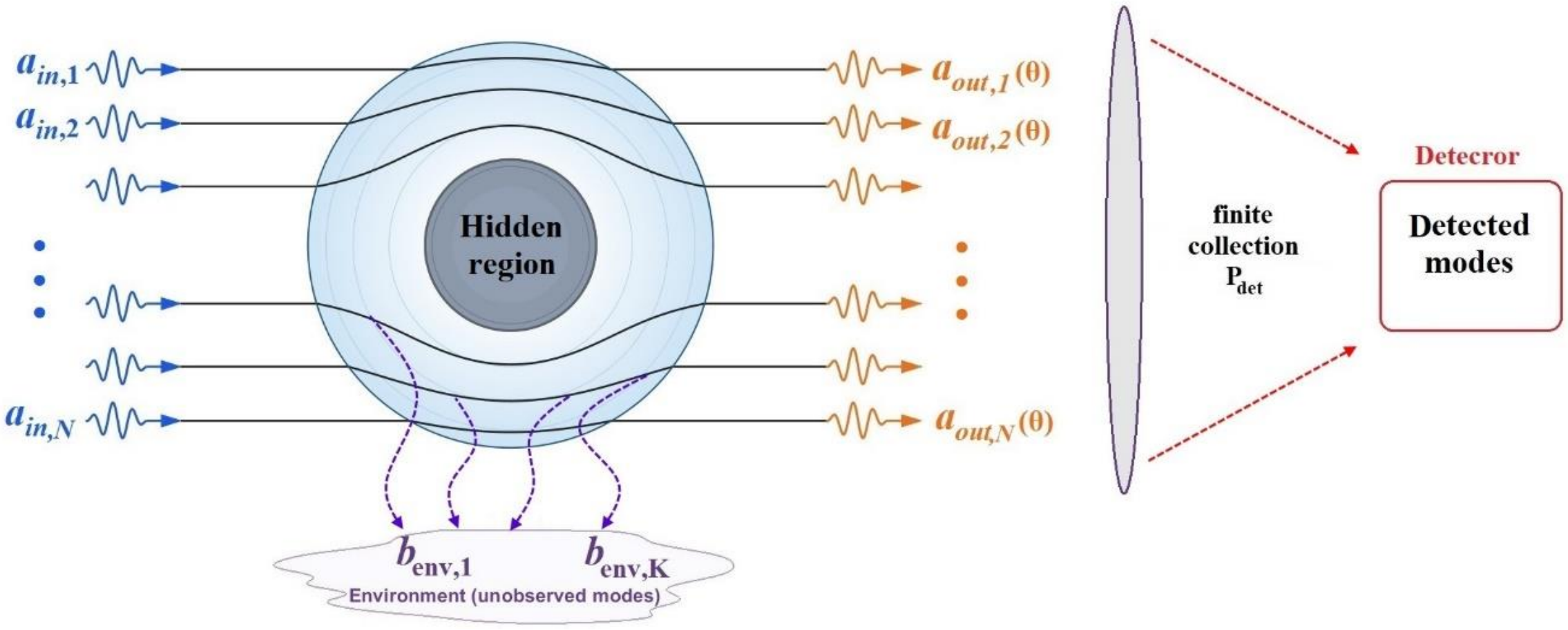

**Figure 1.** Quantum-state detectability framework for a classically cloaked system Incident optical modes $\hat{\mathbf{a}}_{\mathrm{in}}$ interact with a linear passive cloak enclosing a hidden region characterized by the parameter $\theta$. The accessible outgoing modes $\hat{\mathbf{a}}_{\mathrm{out}}(\theta)$ may retain a hidden-object imprint, while absorption, material reservoirs, and uncollected radiation are represented by unobserved environmental modes $\hat{\mathbf{b}}_{\mathrm{env}}$. A finite collection aperture projects the outgoing field onto the detected modal subspace through $P_{\mathrm{det}}$, producing the detected modes measured by the detector.

For Gaussian input states and Gaussian environmental reservoirs, Eq. (6) induces a Gaussian channel on the detected quadrature variables [24–31]. We introduce the quadrature vector

$$\hat{\boldsymbol{\xi}} = (\hat{x}_1, \hat{p}_1, \dots, \hat{x}_N, \hat{p}_N)^T \tag{9}$$

with canonical commutation relations

$$[\hat{\xi}_j, \hat{\xi}_k] = i\Omega_{jk} \tag{10}$$

where Ω is the symplectic form. A Gaussian state is fully characterized by its first moment

$$\boldsymbol{\mu} = \langle \hat{\boldsymbol{\xi}} \rangle \tag{11}$$

and covariance matrix

$$V_{jk} = \frac{1}{2}\langle \Delta\hat{\xi}_j \Delta\hat{\xi}_k + \Delta\hat{\xi}_k \Delta\hat{\xi}_j \rangle \tag{12}$$

The action of the cloaked system on these moments can be written as

$$\boldsymbol{\mu}_{\mathrm{out}}(\theta) = X(\theta)\boldsymbol{\mu}_{\mathrm{in}} + \mathbf{d}(\theta) \tag{13}$$

$$V_{\text{out}}(\theta) = X(\theta)V_{\text{in}}X^T(\theta) + Y(\theta) \qquad (14)$$

Here $X(\theta)$ is the real quadrature representation of the coherent part of the input–output transformation, $Y(\theta)$ is the added-noise matrix associated with loss and inaccessible modes, and $\mathbf{d}(\theta)$ is a possible coherent displacement. For passive scattering without additional coherent emission, $\mathbf{d}(\theta)$ may vanish, but it is retained to keep the channel representation general.

With the convention in Eqs. (10)–(12), a physical Gaussian channel must satisfy the complete-positivity condition

$$Y(\theta) + \frac{i}{2}\Omega - \frac{i}{2}X(\theta)\Omega X^T(\theta) \geq 0 \qquad (15)$$

This condition guarantees that the output covariance matrix satisfies the uncertainty principle for every physical input state. In the lossless fully accessible limit, $Y(\theta) = 0$, and Eq. (15) reduces to the symplectic condition

$$X(\theta)\Omega X^T(\theta) = \Omega \qquad (16)$$

Thus, an ideal lossless cloak corresponds to a canonical transformation on the full optical phase space. In realistic detected channels, however, tracing over unobserved modes produces a reduced noisy Gaussian channel.

A detector measures only a finite set of modes. We denote the corresponding projection onto the accessible detection subspace by $P_{\text{det}}$. The detected first and second moments are therefore

$$\boldsymbol{\mu}_{\text{det}}(\theta) = P_{\text{det}}\boldsymbol{\mu}_{\text{out}}(\theta) \qquad (17)$$
$$V_{\text{det}}(\theta) = P_{\text{det}}V_{\text{out}}(\theta)P_{\text{det}}^T \qquad (18)$$

Combining Eqs. (13), (14), (17), and (18), the operational detected Gaussian channel becomes

$$\boldsymbol{\mu}_{\text{det}}(\theta) = X_{\text{det}}(\theta)\boldsymbol{\mu}_{\text{in}} + \mathbf{d}_{\text{det}}(\theta) \qquad (19)$$

$$V_{\text{det}}(\theta) = X_{\text{det}}(\theta)V_{\text{in}}X_{\text{det}}^T(\theta) + Y_{\text{det}}(\theta) \qquad (20)$$

where

$$X_{\text{det}}(\theta) = P_{\text{det}}X(\theta), \mathbf{d}_{\text{det}}(\theta) = P_{\text{det}}\mathbf{d}(\theta) \qquad (21)$$

and

$$Y_{\text{det}}(\theta) = P_{\text{det}}Y(\theta)P_{\text{det}}^T \qquad (22)$$

Equations (19) and (20) define the detected state used in the rest of the paper. They show that quantum detectability is controlled not by the full electromagnetic field alone, but by the part of the hidden-object imprint that survives in the measured Gaussian moments. A cloaked object is locally undetectable on the detected subspace only if

$$\partial_\theta\boldsymbol{\mu}_{\text{det}}(\theta) = 0, \qquad \partial_\theta V_{\text{det}}(\theta) = 0 \qquad (23)$$

If either derivative in Eq. (23) is nonzero, the detected Gaussian state retains local information about the hidden parameter. The next section quantifies this statement using quantum Fisher information.

## 4. QFI Criterion for Quantum Undetectability

Having reduced the detected output of the cloaked system to the Gaussian state specified by Eqs. (19) and (20), we now introduce the quantum Fisher information criterion for local detectability. The hidden object is represented by the parameter $\theta$. The operational question is whether $\theta$ can be estimated from measurements performed only on the detected output state $\rho_{\mathrm{det}}(\theta)$.

For a parameter-dependent quantum state, the quantum Fisher information $F_Q(\theta)$ sets the ultimate local precision bound through the quantum Cramér–Rao inequality,

$$\mathrm{Var}(\hat{\theta}) \geq \frac{1}{M F_Q(\theta)}, \qquad (24)$$

where $M$ is the number of independent repetitions. Thus, $F_Q(\theta) > 0$ means that the parameter is locally estimable in principle, whereas $F_Q(\theta) = 0$ means that the state carries no local information about that parameter.

For the detected state of a cloaked system, the relevant quantity is therefore

$$F_Q^{\mathrm{det}}(\theta) \equiv F_Q[\rho_{\mathrm{det}}(\theta)]. \qquad (25)$$

Local quantum undetectability on the detected subspace requires

$$F_Q^{\mathrm{det}}(\theta) = 0. \qquad (26)$$

This condition is stronger than the suppression of a selected classical scattering observable. In particular, a small value of $\sigma_{\mathrm{sca}}$ does not by itself imply Eq. (26), because $\sigma_{\mathrm{sca}}$ constrains only a classical field signature rather than the complete detected quantum state.

For Gaussian states, the quantum Fisher information is determined by the parameter dependence of the displacement vector and the covariance matrix. In the notation of Section 3, the detected state is specified by $\boldsymbol{\mu}_{\mathrm{det}}(\theta)$ and $V_{\mathrm{det}}(\theta)$. Therefore, local quantum undetectability requires

$$\partial_\theta \boldsymbol{\mu}_{\mathrm{det}}(\theta) = 0, \partial_\theta V_{\mathrm{det}}(\theta) = 0. \qquad (27)$$

If both conditions in Eq. (27) are satisfied, the detected Gaussian state is locally independent of the hidden parameter. If either condition fails, the detected state retains a local parameter imprint and the corresponding QFI can become nonzero.

The displacement contribution to the Gaussian-state QFI is especially transparent. When the covariance matrix is independent of $\theta$, or when the displacement contribution is isolated from the covariance contribution, the detected displacement contribution is

$$F_\mu^{\mathrm{det}}(\theta) = [\partial_\theta \boldsymbol{\mu}_{\mathrm{det}}(\theta)]^T V_{\mathrm{det}}^{-1} [\partial_\theta \boldsymbol{\mu}_{\mathrm{det}}(\theta)] \qquad (28)$$

Using Eq. (19), the parameter derivative of the detected displacement is

$$\partial_\theta \boldsymbol{\mu}_{\mathrm{det}}(\theta) = [\partial_\theta X_{\mathrm{det}}(\theta)] \boldsymbol{\mu}_{\mathrm{in}} + \partial_\theta \mathbf{d}_{\mathrm{det}}(\theta) \qquad (29)$$

Substitution into Eq. (28) gives

$$F_\mu^{\mathrm{det}}(\theta) = \{[\partial_\theta X_{\mathrm{det}}(\theta)] \boldsymbol{\mu}_{\mathrm{in}} + \partial_\theta \mathbf{d}_{\mathrm{det}}(\theta)\}^T V_{\mathrm{det}}^{-1} \{[\partial_\theta X_{\mathrm{det}}(\theta)] \boldsymbol{\mu}_{\mathrm{in}} + \partial_\theta \mathbf{d}_{\mathrm{det}}(\theta)\} \qquad (30)$$

Equation (30) shows that displacement-level detectability is governed by the derivative of the detected channel with respect to the hidden-object parameter, not by the magnitude of a selected scattering observable alone.

If $V_{\mathrm{det}} \succ 0$, Eq. (28) immediately implies

$$\partial_\theta \boldsymbol{\mu}_{\text{det}}(\theta) \neq 0 \Longrightarrow F_\mu^{\text{det}}(\theta) > 0 \tag{31}$$

This is the displacement-level detectability criterion. Its content is simple but important: any nonzero displacement imprint that survives in the detected Gaussian state is, in principle, locally detectable. Conversely,

$$F_\mu^{\text{det}}(\theta) = 0 \Leftrightarrow \partial_\theta \boldsymbol{\mu}_{\text{det}}(\theta) = 0 \tag{32}$$

provided $V_{\text{det}}$is positive definite.

A useful lower bound follows from the Rayleigh quotient. Defining

$$\mathbf{u}(\theta) = \partial_\theta \boldsymbol{\mu}_{\text{det}}(\theta) \tag{33}$$

one has

$$F_\mu^{\text{det}}(\theta) = \mathbf{u}^T V_{\text{det}}^{-1} \mathbf{u} \geq \frac{\| \mathbf{u} \|^2}{\lambda_{\max}(V_{\text{det}})} \tag{34}$$

Thus,

$$F_\mu^{\text{det}}(\theta) \geq \frac{\| \partial_\theta \boldsymbol{\mu}_{\text{det}}(\theta) \|^2}{\lambda_{\max}(V_{\text{det}})} \tag{35}$$

Equation (35) makes explicit that loss and noise suppress detectability through the detected covariance scale, but they do not eliminate displacement-level detectability unless the detected parameter imprint itself vanishes or is projected outside the accessible subspace.

The covariance contribution provides an independent route to detectability. Even when

$$\partial_\theta \boldsymbol{\mu}_{\text{det}}(\theta) = 0 \tag{36}$$

the detected state may still carry information through

$$\partial_\theta V_{\text{det}}(\theta) \neq 0 \tag{37}$$

In that case, the covariance part of the Gaussian QFI is nonzero whenever the covariance change produces a distinguishable local deformation of the detected state [27–31]. The full Gaussian-state QFI formula is given in Appendix A. In the main text we use only the implication that local Gaussian-state undetectability requires both derivatives in Eq. (27) to vanish.

The central conclusion of this section can therefore be summarized as

$$\sigma_{\text{sca}}(\theta) \to 0 \ \not\Rightarrow \ F_Q^{\text{det}}(\theta) \to 0 \tag{38}$$

Classical scattering suppression constrains a selected field observable, whereas quantum-state undetectability requires the stronger condition that the detected state itself be locally independent of $\theta$. The next section connects this criterion to a physical regularized cylindrical cloak and compares the scaling of classical scattering with derivative-based quantum sensitivity.

## 5. Physical Benchmark: Regularized Cylindrical Cloak

The Gaussian-channel criterion derived above is general: it states that quantum-state undetectability requires the detected state to be locally independent of the hidden-object parameter. However, to make this criterion physically concrete, it is useful to examine a transformation-optical cloak in which the relation between classical scattering suppression and derivative-based quantum sensitivity can be written explicitly. We therefore consider a regularized cylindrical cloak as an analytically tractable benchmark [1–4,12–14].

An ideal cylindrical transformation-optical cloak maps a finite virtual region into a physical annular shell [1–5]. In the singular limit, the inner boundary of the cloak corresponds to a point in the virtual electromagnetic space. If the material parameters are implemented exactly, the cloak is lossless and impedance matched, and the detector has access only to exterior fields, the exterior response can become independent of the object placed inside the cloaked region. In that mathematical limit, the detected state may satisfy

$$\partial_\theta \boldsymbol{\mu}_{\mathrm{det}}(\theta) = 0, \partial_\theta V_{\mathrm{det}}(\theta) = 0 \qquad (39)$$

and the hidden object is locally quantum-undetectable on the detected exterior subspace.

Realistic cloaks, however, are necessarily regularized. Singular material parameters are replaced by finite values, the inner boundary is not compressed to a mathematical point, and the exterior response can retain a residual dependence on the hidden object [12–14]. A simple representation of this regularization is to replace the singular virtual radius by a finite effective radius

$$R = a_{\mathrm{eff}} = \rho a, 0 < \rho < 1 \qquad (40)$$

where $a$is the physical radius of the concealed region and $\rho$ is a dimensionless regularization parameter. The singular-cloak limit is recovered as $\rho \to 0$.

In a scalar cylindrical modal benchmark, the exterior response of the regularized cloak may be represented by the scattering response of an effective virtual cylinder of radius $R = \rho a$ [12–17]. If the hidden-object parameter is chosen as the relative permittivity $\epsilon_{\mathrm{obj}}$, the exterior modal coefficient of angular order $m$can be written schematically as

$$s_m^{\mathrm{cloak}}(\epsilon_{\mathrm{obj}}, \rho) = s_m^{\mathrm{cyl}}(\epsilon_{\mathrm{obj}}, k_0 \rho a) \qquad (41)$$

where $s_m^{\mathrm{cyl}}$is the corresponding cylindrical scattering coefficient in the virtual space. Equation (41) is not intended as a full vectorial simulation of an anisotropic multilayer cloak. It is used as a controlled analytical benchmark that captures the essential effect of regularization: a finite virtual radius allows the exterior field to retain a finite sensitivity to the hidden-object parameter.

To make this sensitivity explicit, consider the weak-contrast Born limit [15–17]. Let

$$\chi = \epsilon_{\mathrm{obj}} - 1 \qquad (42)$$

be the scalar permittivity contrast of the effective virtual cylinder. To first order in $\chi$, the modal scattering coefficient can be written as

$$s_m^{\mathrm{cloak}}(\epsilon_{\mathrm{obj}}, \rho) \simeq C_m(\epsilon_{\mathrm{obj}} - 1) \int_0^{\rho a} r J_m^2(k_0 r)\, dr \qquad (43)$$

where $J_m$ is the Bessel function of the first kind and $C_m$ is a nonzero convention-dependent coefficient that contains the phase and normalization of the cylindrical scattering basis.

The radial integral in Eq. (43) has the exact closed form

$$\int_0^R r J_m^2(k_0 r)\, dr = \frac{R^2}{2}[J_m^2(k_0 R) - J_{m-1}(k_0 R)\, J_{m+1}(k_0 R)] \quad (44)$$

Using $R = \rho a$, Eq. (43) becomes

$$s_m^{\text{cloak}}(\epsilon_{\text{obj}}, \rho) \simeq C_m(\epsilon_{\text{obj}} - 1) \times \frac{(\rho a)^2}{2}[J_m^2(k_0 \rho a) - J_{m-1}(k_0 \rho a) J_{m+1}(k_0 \rho a)] \quad (45)$$

Differentiating with respect to the hidden-object permittivity gives

$$\partial_{\epsilon_{\text{obj}}} s_m^{\text{cloak}}(\epsilon_{\text{obj}}, \rho) \simeq C_m \int_0^{\rho a} r J_m^2(k_0 r)\, dr \quad (46)$$

Since $r J_m^2(k_0 r) \geq 0$ and is not identically zero on any finite interval $0 < r < \rho a$, one obtains

$$\partial_{\epsilon_{\text{obj}}} s_m^{\text{cloak}}(\epsilon_{\text{obj}}, \rho) \neq 0 \text{for} \rho > 0 \quad (47)$$

Thus, any finite regularization radius leaves a nonzero local dependence of the exterior modal coefficient on the hidden-object permittivity within this benchmark model.

For $k_0 \rho a \ll 1$, the small-argument expansion

$$J_m(x) \simeq \frac{1}{m!}\left(\frac{x}{2}\right)^m \quad (48)$$

gives

$$\int_0^{\rho a} r J_m^2(k_0 r)\, dr \simeq \frac{k_0^{2m} (\rho a)^{2m+2}}{(2m+2) 2^{2m} (m!)^2} \quad (49)$$

Therefore,

$$s_m^{\text{cloak}} \simeq C_m(\epsilon_{\text{obj}} - 1) \frac{k_0^{2m} (\rho a)^{2m+2}}{(2m+2) 2^{2m} (m!)^2} \quad (50)$$

and

$$\partial_{\epsilon_{\text{obj}}} s_m^{\text{cloak}} \simeq C_m \frac{k_0^{2m} (\rho a)^{2m+2}}{(2m+2) 2^{2m} (m!)^2} \quad (51)$$

For the dominant monopole term $m = 0$, these reduce to

$$s_0^{\text{cloak}} \simeq C_0(\epsilon_{\text{obj}} - 1) \frac{(\rho a)^2}{2} \quad (52)$$

and

$$\partial_{\epsilon_{\text{obj}}} s_0^{\text{cloak}} \simeq C_0 \frac{(\rho a)^2}{2} \quad (53)$$

Equations (52) and (53) show two important features. First, for every finite $\rho > 0$, the derivative with respect to the hidden-object parameter is nonzero. Second, both the modal coefficient and its derivative vanish algebraically as $\rho \to 0$, consistent with recovery of the ideal singular-cloak limit.

The distinction between classical scattering strength and quantum-state sensitivity is now explicit. A classical scattering observable is controlled by the modal intensity sum [15–17]

$$\sigma_{\text{sca}}^{\text{cloak}} \propto \sum_m w_m \mid s_m^{\text{cloak}} \mid^2 \quad (54)$$

where $w_0 = 1$ and $w_m = 2$ for $m > 0$ account for the $\pm m$ modal degeneracy. By contrast, the displacement-level quantum sensitivity in a detected modal subspace is controlled by

$$F_{\mu}^{\mathrm{det,cloak}} \propto \sum_{m \in \mathrm{det}} w_m \mid \partial_{\epsilon_{\mathrm{obj}}} s_m^{\mathrm{cloak}} \mid^2, \quad (55)$$

up to the detected quadrature-noise normalization discussed in Section 4. Equations (54) and (55) are related, but they are not identical. The classical scattering signal depends on the magnitude of the residual modal amplitude, whereas the QFI contribution depends on the local derivative of the detected state with respect to the hidden-object parameter.

This distinction is especially clear near a reference value where the scattering amplitude itself vanishes. For example, at $\epsilon_{\mathrm{obj}} = 1$, the Born-level contrast is zero and hence

$$s_m^{\mathrm{cloak}} = 0. \quad (56)$$

Nevertheless, for every finite regularization parameter $\rho > 0$,

$$\partial_{\epsilon_{\mathrm{obj}}} s_m^{\mathrm{cloak}} \simeq C_m \int_0^{\rho a} r\, J_m^2(k_0 r)\, dr \neq 0 \quad (57)$$

Thus, vanishing of a selected scattering amplitude at a reference parameter value does not imply vanishing local sensitivity to changes in that parameter. This distinction is illustrated in Fig. 2 using the dominant monopole Born term. From Eq. (52), the scattering coefficient satisfies $s_0^{\mathrm{cloak}} \propto (\epsilon_{\mathrm{obj}} - 1) I_0(\rho)$, where $I_0(\rho)$ is the monopole Bessel overlap integral. Hence the classical scattering strength scales as $\left|s_0^{\mathrm{cloak}}\right|^2 \propto (\epsilon_{\mathrm{obj}} - 1)^2 |I_0(\rho)|^2$ and vanishes at the reference contrast $\epsilon_{\mathrm{obj}} = 1$. By contrast, Eq. (53) gives $\partial_{\epsilon_{\mathrm{obj}}} s_0^{\mathrm{cloak}} \propto I_0(\rho)$, so the derivative sensitivity $\left|\partial_{\epsilon_{\mathrm{obj}}} s_0^{\mathrm{cloak}}\right|^2$ remains finite for every finite $\rho > 0$. The QFI captures precisely this derivative-based distinguishability.

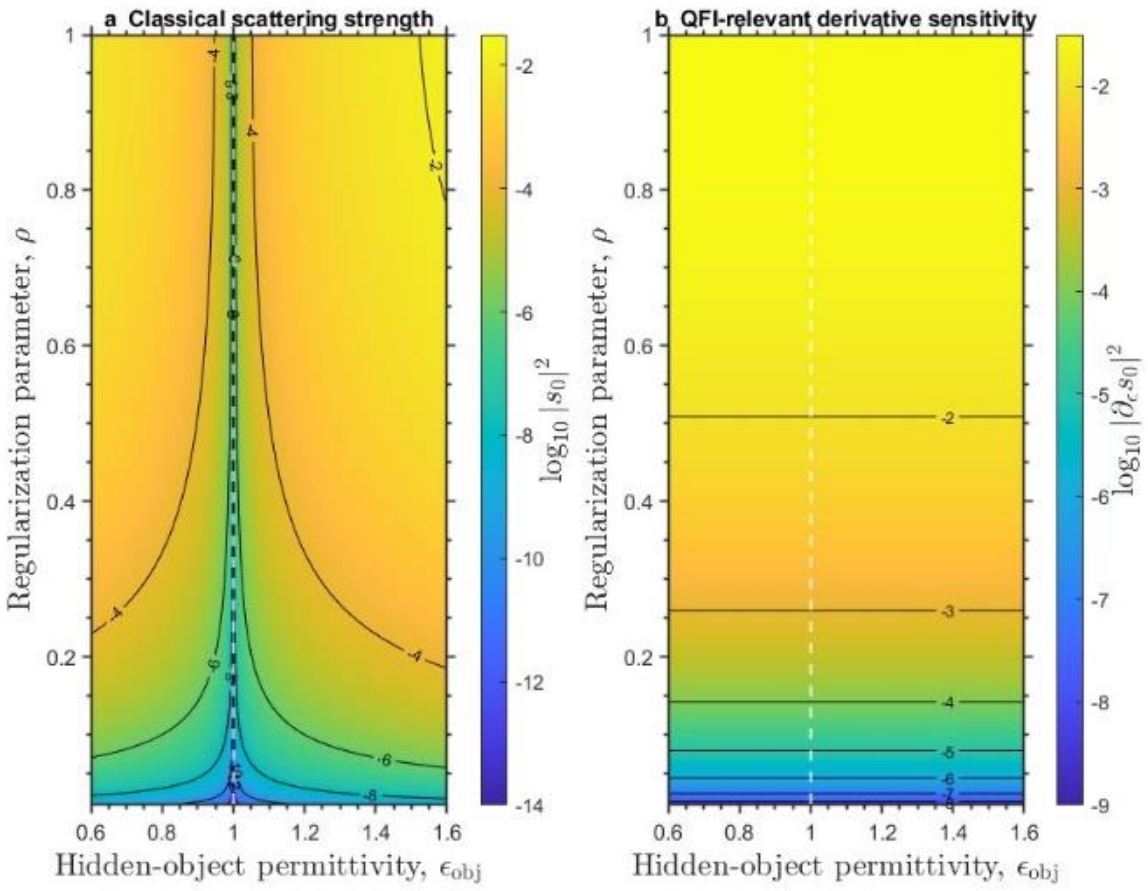

**Figure 2.** Regularized cylindrical cloak benchmark in the Born limit. a, Two-parameter map of the Born-level classical scattering strength $\log_{10} \mid s_0 \mid^2$, where $s_0 \equiv s_0^{\mathrm{cloak}}$, evaluated from the exact monopole Bessel overlap integral. The vertical dashed line marks the reference value $\epsilon_{\mathrm{obj}} = 1$, where the contrast vanishes and the scattering amplitude is zero. b, Corresponding map of the derivative sensitivity $\log_{10} \mid \partial_\epsilon s_0 \mid^2$, with $\epsilon \equiv \epsilon_{\mathrm{obj}}$, which enters the displacement-level QFI. Unlike the scattering strength, the derivative sensitivity does not vanish at $\epsilon_{\mathrm{obj}} = 1$ for finite $\rho > 0$. The benchmark illustrates that suppressing a scattering amplitude at a reference parameter value is not equivalent to eliminating local quantum-state sensitivity.

The regularized cylindrical cloak benchmark therefore provides a concrete interpretation of the general criterion derived in Section 4. In the ideal singular limit, $\rho \to 0$, the exterior modal response becomes increasingly insensitive to the hidden object, and quantum-state undetectability may be recovered on the exterior detected subspace. For any finite regularization, however, the exterior modal coefficients retain a reduced but nonzero local sensitivity to the hidden-object parameter. Whether this sensitivity is operationally accessible depends on loss, noise, and the modal subspace collected by the detector, which we analyze next.

## 6. Loss, Noise, and Finite Aperture

The benchmark of Section 5 shows that a finite regularization parameter can leave a derivative-level imprint of the hidden object in the exterior modal response. Whether this imprint is operationally detectable, however, depends on how it reaches the measured modes. Material absorption, environmental noise, imperfect mode overlap, and finite numerical aperture do not merely reduce an intensity signal; they define the reduced Gaussian channel on which the detected state is measured [18–24]. In this section we separate two physically distinct effects. First, loss and environmental noise reduce the precision with which a surviving imprint can be estimated. Second, finite numerical aperture projects the outgoing field onto a restricted modal subspace and can remove the imprint from the detector altogether [38–41].

We first consider a single effective detected mode in order to isolate the role of loss and environmental noise. Let $\eta$denote an effective transmissivity that includes propagation through the cloak, collection efficiency, and mode overlap with the detector. For a displacement-encoded hidden-parameter imprint, we write

$$\mu_{\mathrm{det}}(\theta) = \sqrt{\eta}\,\mu_{\mathrm{in}}(\theta) \qquad (58)$$

and

$$V_{\mathrm{det}} = \eta V_{\mathrm{in}} + (1-\eta)V_{\mathrm{env}} \qquad (59)$$

where $V_{\mathrm{env}}$ is the quadrature variance of the environmental mode coupled into the detected channel. In this minimal benchmark, $0 \le \eta \le 1$. The case $\eta = 1$ corresponds to an ideal detected channel with no environmental admixture, whereas $\eta = 0$ corresponds to complete loss of the detected displacement imprint.

Using the displacement contribution to the Gaussian QFI, the detected displacement-level QFI becomes

$$F_\mu^{\mathrm{det}}(\theta) = \frac{\eta(\partial_\theta \mu_{\mathrm{in}})^2}{\eta V_{\mathrm{in}} + (1-\eta)V_{\mathrm{env}}} \qquad (60)$$

The corresponding freely accessible value is

$$F_\mu^{\mathrm{free}}(\theta) = \frac{(\partial_\theta \mu_{\mathrm{in}})^2}{V_{\mathrm{in}}} \qquad (61)$$

Therefore,

$$\frac{F_\mu^{\mathrm{det}}}{F_\mu^{\mathrm{free}}} = \frac{\eta V_{\mathrm{in}}}{\eta V_{\mathrm{in}} + (1-\eta)V_{\mathrm{env}}} \qquad (62)$$

Equation (62) is the closed-form expression used in Fig. 3. It shows that loss and environmental noise reduce the accessible QFI, but do not force it to vanish for finite $\eta > 0$, finite $V_{\mathrm{env}}$, and nonzero $\partial_\theta \mu_{\mathrm{in}}$. Instead, vanishing displacement-level detectability in this benchmark requires either complete removal of the detected channel or removal of the displacement imprint itself,

$$F_\mu^{\mathrm{det}} = 0 \Leftrightarrow \eta = 0 \text{ or } \partial_\theta \mu_{\mathrm{in}} = 0 \qquad (63)$$

assuming finite $V_{\text{env}}$. This result is consistent with the general QFI criterion derived above: finite noise increases the covariance scale and degrades precision, but it does not certify quantum-state undetectability unless the detected derivative itself vanishes.

The same conclusion can be connected to the regularized cloak benchmark. If the detected displacement imprint is proportional to the exterior modal derivative $\partial_{\epsilon_{\text{obj}}} s_m^{\text{cloak}}$, then the detected QFI in the lossy single-mode approximation scales as

$$F_{\mu,m}^{\text{det}} \propto \frac{\eta \mid \partial_{\epsilon_{\text{obj}}} s_m^{\text{cloak}} \mid^2}{\eta V_{\text{in}} + (1-\eta) V_{\text{env}}} \qquad (64)$$

Thus, cloak regularization controls the numerator through the surviving modal sensitivity, whereas loss and environmental noise control the denominator through the detected covariance. These are physically distinct mechanisms: cloak design may reduce the hidden-object imprint, while loss and noise reduce the measurement precision associated with whatever imprint remains.

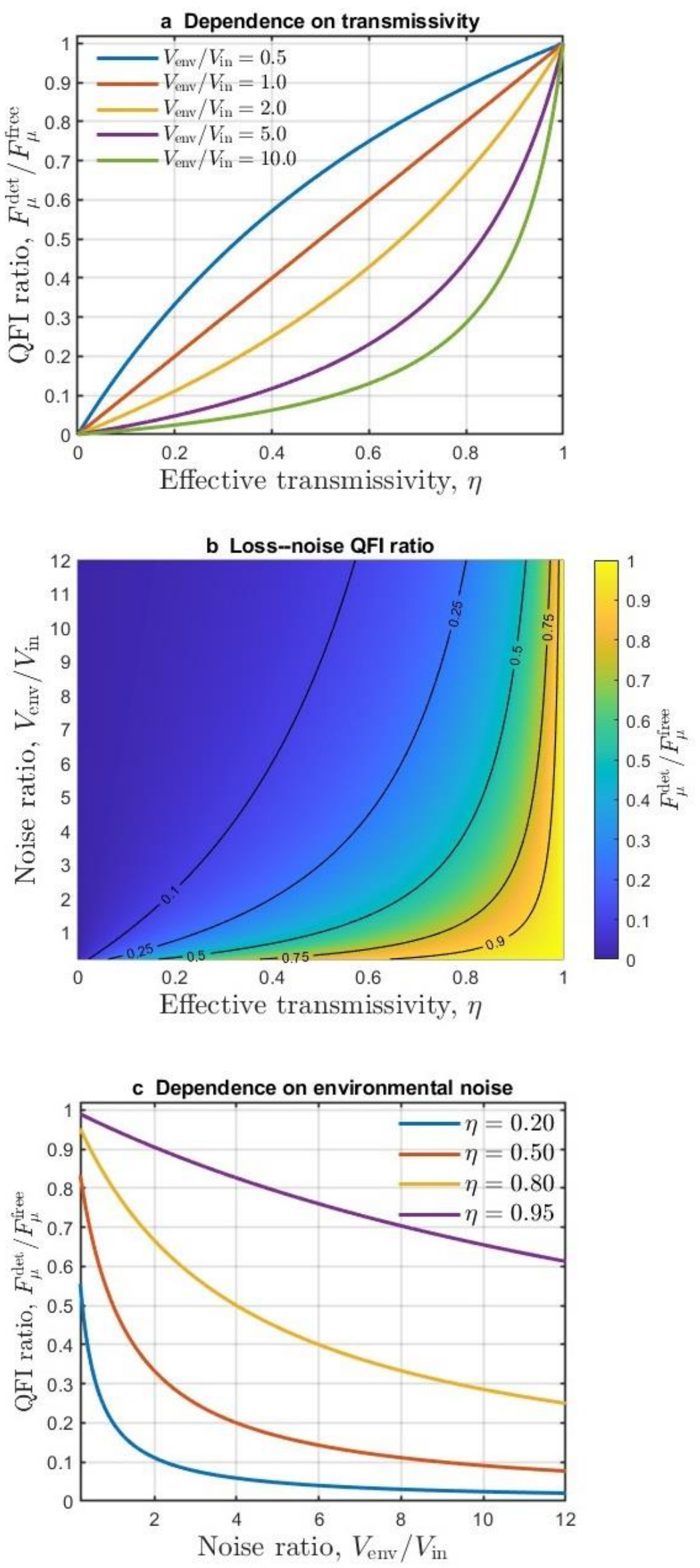

Figure 3. Loss- and noise-limited displacement-level quantum Fisher information. a, QFI ratio $F_\mu^{\text{det}}/F_\mu^{\text{free}}$ as a function of the effective transmissivity $\eta$for different environmental-to-input noise ratios $V_{\text{env}}/V_{\text{in}}$. b, Two-parameter map of the same QFI ratio in the $(\eta, V_{\text{env}}/V_{\text{in}})$ plane. c, Dependence of the QFI ratio on environmental noise for representative fixed values of $\eta$. The benchmark follows from the single-mode Gaussian attenuation model $V_{\text{det}} = \eta V_{\text{in}} + (1-\eta) V_{\text{env}}$. Loss and environmental noise degrade the accessible QFI, but for finite noise they do not eliminate displacement-level detectability unless the detected channel is extinguished or the parameter imprint itself is absent.

Finite numerical aperture introduces a different limitation. Unlike loss and environmental noise, which reduce the precision associated with a surviving imprint, finite aperture projects the outgoing field onto an accessible modal subspace. As a result, the full outgoing field may contain hidden-object information even when the detected subspace does not.

Let

$$\mathbf{u}_{\text{out}}(\theta) = \partial_\theta \boldsymbol{\mu}_{\text{out}}(\theta) \quad (65)$$

be the displacement-level imprint in the full outgoing modal space. The detected imprint is

$$\mathbf{u}_{\text{det}}(\theta) = P_{\text{det}} \mathbf{u}_{\text{out}}(\theta). \quad (66)$$

Substitution into the displacement-level QFI bound gives

$$F_\mu^{\text{det}}(\theta) \geq \frac{\| P_{\text{det}} \, \partial_\theta \boldsymbol{\mu}_{\text{out}}(\theta) \|^2}{\lambda_{\max}(V_{\text{det}})} \quad (67)$$

Equation (67) shows that finite aperture limits detectability by selecting the part of the imprint that lies inside the measured subspace. If

$$P_{\text{det}} \, \partial_\theta \boldsymbol{\mu}_{\text{out}}(\theta) = 0, \quad (68)$$

then the displacement-level QFI vanishes on the detected subspace even if the full outgoing field contains parameter-dependent information. Conversely, if the projected imprint is nonzero and the detected covariance is finite, then the accessible displacement-level QFI is nonzero.

To illustrate this projection effect, consider an angular imprint $| Z(u; u_c) |^2$ in the normalized transverse-wavevector coordinate

$$u = \frac{k_x}{k_0}, -1 \leq u \leq 1. \quad (69)$$

A detector with numerical aperture NA collects only the angular interval $| u | \leq \text{NA}$. The retained imprint fraction is then

$$\mathcal{B}(\text{NA}, u_c) = \frac{\int_{-\text{NA}}^{\text{NA}} | Z(u; u_c) |^2 \, du}{\int_{-1}^{1} | Z(u; u_c) |^2 \, du} \quad (70)$$

Figure 4 shows this retained fraction for a representative Gaussian angular imprint. The purpose of this model is not to prescribe a universal angular spectrum, but to illustrate the operational meaning of the projection $P_{\text{det}}$. A detector with finite numerical aperture accesses only the part of the parameter imprint lying inside its angular collection window. Thus, finite aperture is not equivalent to a universal scalar attenuation factor; it is a projection onto the accessible angular subspace.

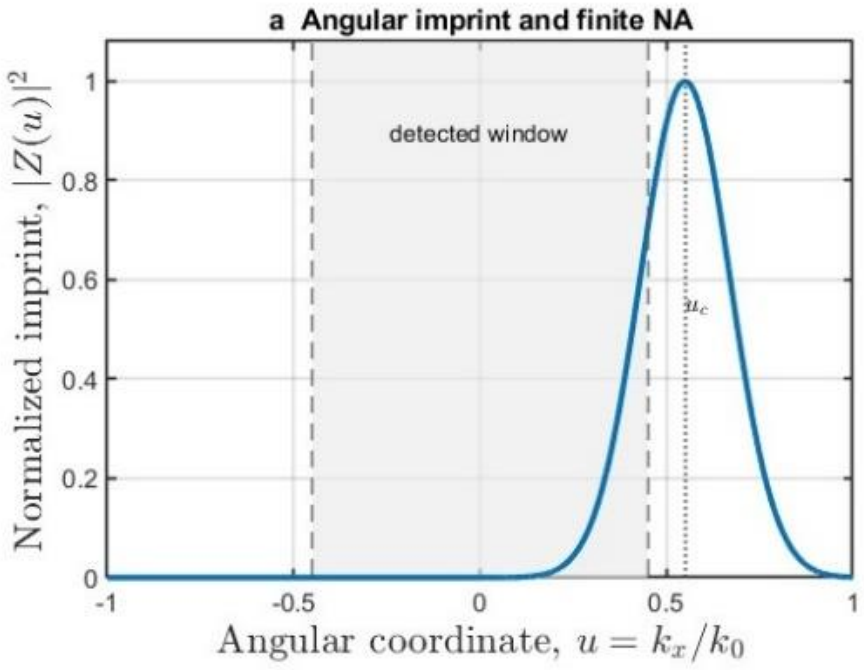

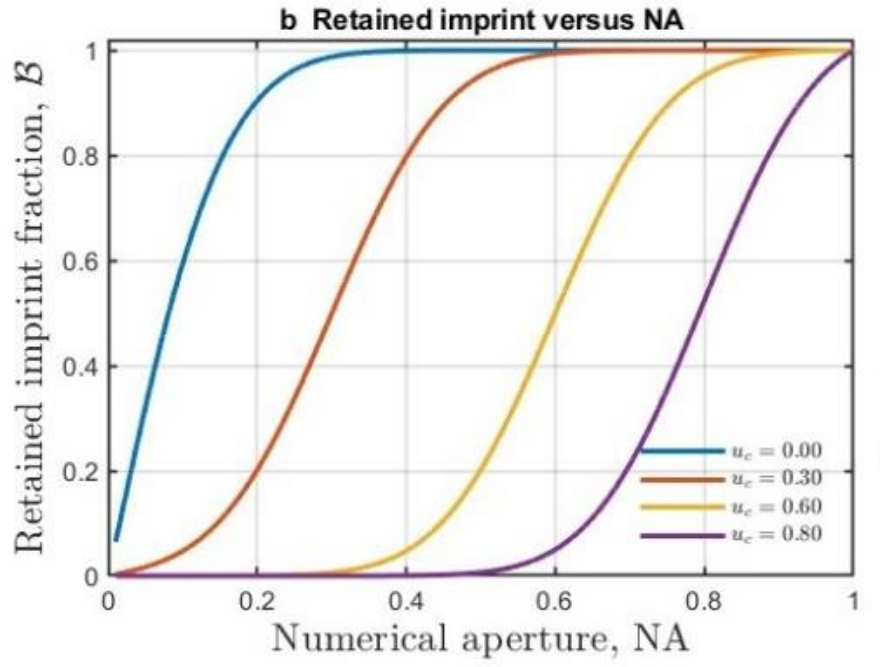

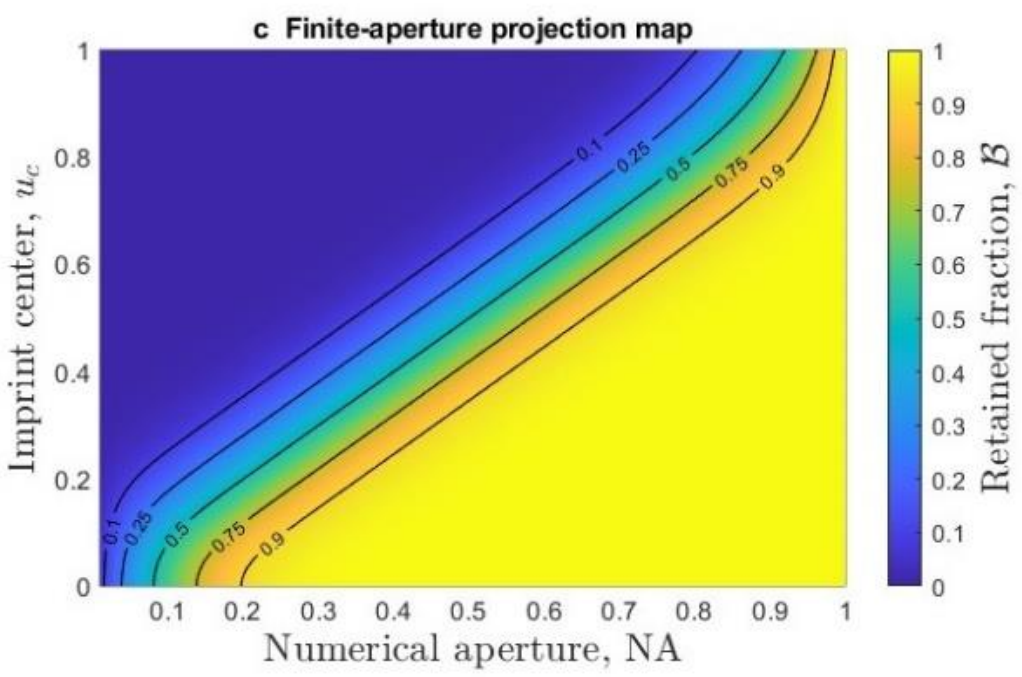

Figure 4. Finite-aperture projection of the hidden-object imprint. a, Representative angular imprint $|Z(u)|^2$, with $u = k_x/k_0$, and the finite numerical-aperture detection window $|u| \leq \mathrm{NA}$. In the displayed example, the imprint center is located near the edge of the detected angular region, so only part of the imprint is collected. b, Retained imprint fraction $\mathcal{B}(\mathrm{NA})$ as a function of numerical aperture for different imprint centers $u_c$. Imprints centered near the optical axis are collected more efficiently than imprints displaced toward high-angle modes. c, Two-parameter map of $\mathcal{B}(\mathrm{NA}, u_c)$, showing that finite aperture acts as a projection onto the accessible angular subspace rather than as a universal scalar attenuation factor. The accessible QFI can vanish on the detected subspace when $P_{\mathrm{det}}\, \partial_\theta \boldsymbol{\mu}_{\mathrm{out}} = 0$, even if the full outgoing field retains hidden-object information.

For the regularized cylindrical cloak, the projection $P_{\mathrm{det}}$ may select only a finite set of angular, spatial, polarization, or frequency modes. In a modal representation, Eq. (67) corresponds schematically to

$$F_\mu^{\mathrm{det,cloak}} \gtrsim \frac{\sum_{m\in\mathrm{det}} w_m \,|\, \partial_{\epsilon_{\mathrm{obj}}} s_m^{\mathrm{cloak}} \,|^2}{\lambda_{\max}(V_{\mathrm{det}})} \quad (71)$$

This expression makes the physical role of finite aperture explicit. A cloak may redistribute the hidden-object imprint into angular or modal components that are poorly collected by the detector. In that case, the accessible QFI is strongly reduced. However, quantum-state undetectability is reached only when the projected derivative in the detected subspace vanishes, not merely when the total scattered intensity is small.

Equations (62), (67), and (71) summarize the physical content of this section. Loss and environmental noise increase the effective covariance and reduce the precision with which a surviving imprint can be estimated. Finite numerical aperture projects the imprint onto the measured subspace and may remove it completely from the detector. Therefore, realistic imperfections degrade quantum detectability, but they do not make classical scattering suppression equivalent to quantum-state undetectability. The latter requires the stronger condition that the hidden-object parameter leave no displacement or covariance imprint in the detected Gaussian state.

## 7. Discussion

The results of this work distinguish two notions that are often treated as equivalent: classical cloaking and quantum-state undetectability. Classical transformation-optical cloaks are usually designed to suppress selected scattering observables, such as far-field amplitudes, cross sections, or exterior field distortions. These are observable-level criteria. Quantum-state undetectability is stronger: the detected optical state itself must be independent of the hidden-object parameter. For Gaussian detected states, this requires both the detected displacement vector and covariance matrix to be locally independent of that parameter.

This distinction does not imply that classical cloaking is ineffective. In an ideal singular transformation-optical limit, the exterior parameter imprint may be removed from the accessible field, and the detected quantum state may then become independent of the hidden object. The point is more specific: suppression of a selected classical scattering amplitude is not sufficient, by itself, to certify quantum-state undetectability. The relevant operational quantity is the derivative of the detected state with respect to the hidden parameter. Whenever a nonzero parameter imprint survives in the detected displacement or covariance, the corresponding quantum Fisher information can remain nonzero.

The regularized cylindrical benchmark makes this statement explicit. In the Born-limit model, the scattering amplitude can be made small by reducing the effective mapped radius and can vanish at a reference contrast. Nevertheless, for any finite regularization parameter, the derivative of the modal coefficient with respect to the hidden-object parameter remains finite. Thus, local detectability is governed not by the magnitude of the scattering amplitude alone, but by the parameter derivative of the detected state. Loss, environmental noise, and finite aperture then determine how much of this derivative-level imprint remains operationally accessible. Loss and noise degrade estimation precision through the detected covariance, whereas finite aperture projects the imprint onto a restricted modal subspace.

The present analysis is deliberately formulated at the Gaussian-channel level. This makes the framework applicable to linear passive cloaking systems and allows the detectability condition to be expressed directly in terms of detected first and second moments. At the same time, it defines the scope of the result. The cylindrical cloak calculation is a scalar Born-limit benchmark rather than a full-vector simulation of a realistic multilayer cloak. Device-level predictions would require the full electromagnetic scattering problem, including geometry, polarization, material dispersion, loss, illumination, and detector aperture. Similarly, non-Gaussian probes or photon-counting measurements may reveal forms of distinguishability beyond the Gaussian first- and second-moment description considered here.

The framework also suggests a natural interpretation of quantum-limited anti-cloaking. Rather than searching only for residual scattered intensity, an optimal detector should maximize overlap with the parameter derivative of the detected state. In modal language, the relevant modes are those carrying $\partial_\theta \boldsymbol{\mu}_{\mathrm{det}}$ or $\partial_\theta V_{\mathrm{det}}$, not necessarily those with the largest field amplitude. Thus, the central question is not merely whether a scattered field is present, but whether the detected quantum state changes with the hidden parameter.

Overall, classical invisibility and quantum-state undetectability are related but inequivalent. Classical cloaking can strongly suppress observable scattering signatures, and in ideal limits it may remove the detected parameter imprint entirely. For realistic regularized, lossy, and finite-aperture systems, however, undetectability must be tested at the level of the detected quantum state. A cloaked object is locally quantum-state undetectable only when the hidden-object parameter leaves no accessible displacement or covariance imprint in the detected Gaussian state.

## 8. Conclusion

We have formulated the detectability of classically cloaked objects as a quantum-state distinguishability problem. The central result is that suppression of a selected classical scattering observable is not, by itself, sufficient to establish quantum-state undetectability. The relevant operational condition is stronger: the detected optical state must be independent of the hidden-object parameter. For Gaussian detected states, this requires both the detected displacement vector and covariance matrix to carry no local parameter imprint.

Within the effective Gaussian-channel description, this condition is quantified by the detected quantum Fisher information $F_Q^{\mathrm{det}}$. If either $\partial_\theta \boldsymbol{\mu}_{\mathrm{det}}$ or $\partial_\theta V_{\mathrm{det}}$ is nonzero, the detected state can retain local information about the hidden object. The regularized cylindrical cloak benchmark illustrates this distinction explicitly: a classical scattering amplitude may be strongly suppressed or vanish at a reference contrast, while its derivative with respect to the hidden-object parameter remains finite for any finite regularization.

Loss, environmental noise, and finite numerical aperture reduce the operational accessibility of this information. Loss and noise degrade the QFI through attenuation and added covariance, whereas finite aperture projects the parameter imprint onto a restricted detected subspace. These effects can strongly suppress the accessible QFI, but they do not make scattering suppression equivalent to quantum-state undetectability. True local undetectability requires the hidden-object parameter to leave no accessible displacement or covariance imprint in the detected state.

The framework developed here provides a state-level criterion for assessing cloaking beyond classical scattering metrics. It connects transformation-optical cloaking, Gaussian quantum channels, and quantum parameter estimation, and it identifies the relevant object for quantum-limited anti-cloaking: the parameter derivative of the detected optical state. This perspective can be extended to full-vector cloak simulations, dispersive and lossy macroscopic-QED models, non-Gaussian probes, and optimized detection strategies for probing the quantum limits of invisibility.

## Appendix A. Gaussian-State Quantum Fisher Information

This appendix summarizes the Gaussian-state quantum Fisher information used in the main text and fixes the covariance convention employed throughout the paper. We use quadrature operators

$$\hat{\boldsymbol{\xi}} = (\hat{x}_1, \hat{p}_1, \dots, \hat{x}_N, \hat{p}_N)^T \qquad \text{(A1)}$$

satisfying

$$[\hat{\xi}_j, \hat{\xi}_k] = i\Omega_{jk} \qquad \text{(A2)}$$

where $\Omega$ is the symplectic form. With this convention, the covariance matrix is defined as

$$V_{jk} = \frac{1}{2}\langle \Delta\hat{\xi}_j \Delta\hat{\xi}_k + \Delta\hat{\xi}_k \Delta\hat{\xi}_j \rangle \qquad \text{(A3)}$$

and physical Gaussian states satisfy the uncertainty relation

$$V + \frac{i}{2}\Omega \geq 0 \qquad \text{(A4)}$$

A Gaussian state $\rho(\theta)$ is fully specified by its first moment

$$\boldsymbol{\mu}(\theta) = \langle \hat{\boldsymbol{\xi}} \rangle \qquad \text{(A5)}$$

and covariance matrix $V(\theta)$. The quantum Fisher information is defined through the symmetric logarithmic derivative $L_\theta$,

$$\partial_\theta \rho(\theta) = \frac{1}{2}[\rho(\theta)L_\theta + L_\theta \rho(\theta)] \qquad \text{(A6)}$$

as

$$F_Q(\theta) = \mathrm{Tr}[\rho(\theta)L_\theta^2] \qquad \text{(A7)}$$

For a nonsingular mixed Gaussian state, the QFI can be written as the sum of a displacement contribution and a covariance contribution,

$$F_Q(\theta) = F_\mu(\theta) + F_V(\theta) \qquad \text{(A8)}$$

With the convention in Eqs. (A1)–(A4), the displacement contribution is

$$F_\mu(\theta) = [\partial_\theta \boldsymbol{\mu}(\theta)]^T V^{-1}(\theta)[\partial_\theta \boldsymbol{\mu}(\theta)] \qquad \text{(A9)}$$

The covariance contribution can be written in vectorized form as

$$F_V(\theta) = \frac{1}{2}\text{vec}[\partial_\theta V(\theta)]^T \mathcal{M}^{-1} \text{vec}[\partial_\theta V(\theta)] \qquad \text{(A10)}$$

where

$$\mathcal{M} = V(\theta) \otimes V(\theta) - \frac{1}{4}\Omega \otimes \Omega \qquad \text{(A11)}$$

For states at the boundary of the uncertainty relation, Eq. (A10) is understood with the inverse taken on the support of $\mathcal{M}$, equivalently by using the Moore–Penrose inverse. This technical qualification does not affect the local detectability conditions used in the main text.

Equations (A9)–(A11) show that a Gaussian state can carry information about $\theta$through either its first moment or its covariance matrix. If

$$\partial_\theta \boldsymbol{\mu}(\theta) \neq 0 \qquad \text{(A12)}$$

and $V(\theta) \succ 0$, then the displacement contribution is strictly positive,

$$F_\mu(\theta) > 0 \qquad \text{(A13)}$$

Similarly, if

$$\partial_\theta V(\theta) \neq 0 \qquad \text{(A14)}$$

then the covariance contribution can be nonzero, provided the covariance change corresponds to a distinguishable deformation of the Gaussian state. Therefore, a Gaussian state is locally independent of the parameter only if both its first and second moments are locally independent of that parameter,

$$\partial_\theta \boldsymbol{\mu}(\theta) = 0, \partial_\theta V(\theta) = 0 \qquad \text{(A15)}$$

For the detected state of the cloaked system, this condition becomes

$$\partial_\theta \boldsymbol{\mu}_{\text{det}}(\theta) = 0, \partial_\theta V_{\text{det}}(\theta) = 0 \qquad \text{(A16)}$$

Consequently, local quantum-state undetectability on the detected subspace requires

$$F_Q^{\text{det}}(\theta) = 0 \qquad \text{(A17)}$$

which, for Gaussian detected states, is ensured when Eq. (A16) holds. Conversely, if either

$$\partial_\theta \boldsymbol{\mu}_{\text{det}}(\theta) \neq 0 \qquad \text{(A18)}$$

or

$$\partial_\theta V_{\text{det}}(\theta) \neq 0 \qquad \text{(A19)}$$

then the detected Gaussian state generally retains local information about the hidden parameter, and the corresponding quantum Fisher information can be nonzero.

This is the state-level condition used throughout the main text. It is stronger than the suppression of a selected classical scattering observable, because it requires the complete detected Gaussian state, not merely a chosen field amplitude or intensity signature, to be locally independent of the hidden-object parameter.

**Appendix B. Born-Limit Derivation for a Regularized Cylindrical Cloak**

This appendix provides the Born-limit derivation used in Section 5 for the regularized cylindrical cloak benchmark. The purpose is to justify the scaling relation between the classical modal scattering coefficient and the derivative-based quantum sensitivity.

We consider a scalar two-dimensional cylindrical scattering problem in the virtual space associated with the regularized cloak. The singular ideal cloak is replaced by an effective virtual cylinder of radius

$$R = \rho a \qquad \text{(B1)}$$

where $a$ is the physical radius of the concealed region and $0 < \rho < 1$ is the regularization parameter. The hidden-object parameter is taken to be the relative permittivity $\epsilon_{\mathrm{obj}}$, and the corresponding weak contrast is

$$\chi = \epsilon_{\mathrm{obj}} - 1 \qquad \text{(B2)}$$

In the first Born approximation, the scattered field is linear in the contrast $\chi$. Expanding the incident and scattered fields in cylindrical harmonics, the modal coefficient of angular order $m$ is proportional to the overlap between the perturbation and the squared radial cylindrical basis function. We therefore write

$$s_m^{\mathrm{cloak}}(\epsilon_{\mathrm{obj}}, \rho) \simeq C_m \chi \int_0^R r\, J_m^2(k_0 r)\, dr \qquad \text{(B3)}$$

where $J_m$ is the Bessel function of the first kind, $k_0$is the background wavenumber, and $C_m$ is a nonzero coefficient that depends on the normalization convention of the cylindrical scattering basis. Substituting Eq. (B2) into Eq. (B3) gives

$$s_m^{\mathrm{cloak}}(\epsilon_{\mathrm{obj}}, \rho) \simeq C_m (\epsilon_{\mathrm{obj}} - 1) \int_0^R r\, J_m^2(k_0 r)\, dr \qquad \text{(B4)}$$

The radial integral can be evaluated using the standard identity

$$\int xJ_m^2(x)\,dx = \frac{x^2}{2}[J_m^2(x) - J_{m-1}(x)J_{m+1}(x)] \qquad \text{(B5)}$$

Setting $x = k_0 r$, one obtains

$$\int_0^R r\,J_m^2(k_0 r)\,dr = \frac{R^2}{2}[J_m^2(k_0 R) - J_{m-1}(k_0 R)J_{m+1}(k_0 R)] \qquad \text{(B6)}$$

Using $R = \rho a$, Eq. (B4) becomes

$$s_m^{\text{cloak}}(\epsilon_{\text{obj}}, \rho) \simeq C_m(\epsilon_{\text{obj}} - 1)\frac{(\rho a)^2}{2}[J_m^2(k_0\rho a) - J_{m-1}(k_0\rho a)J_{m+1}(k_0\rho a)] \qquad \text{(B7)}$$

Differentiating Eq. (B4) with respect to the hidden-object permittivity gives

$$\partial_{\epsilon_{\text{obj}}} s_m^{\text{cloak}}(\epsilon_{\text{obj}}, \rho) \simeq C_m \int_0^{\rho a} r\,J_m^2(k_0 r)\,dr \qquad \text{(B8)}$$

Since $rJ_m^2(k_0 r) \geq 0$ and is not identically zero on any finite interval $0 < r < \rho a$, Eq. (B8) implies

$$\partial_{\epsilon_{\text{obj}}} s_m^{\text{cloak}}(\epsilon_{\text{obj}}, \rho) \neq 0, \rho > 0 \qquad \text{(B9)}$$

provided $C_m \neq 0$ for the detected scattering channel. Thus, for every finite regularization parameter, the Born-limit modal coefficient retains local sensitivity to the hidden-object permittivity.

For $k_0\rho a \ll 1$, the Bessel function has the small-argument expansion

$$J_m(x) \simeq \frac{1}{m!}\left(\frac{x}{2}\right)^m \qquad \text{(B10)}$$

Therefore,

$$J_m^2(k_0 r) \simeq \frac{k_0^{2m} r^{2m}}{2^{2m}(m!)^2} \qquad \text{(B11)}$$

Substitution into the overlap integral gives

$$\int_0^{\rho a} r\,J_m^2(k_0 r)\,dr \simeq \frac{k_0^{2m}}{2^{2m}(m!)^2}\int_0^{\rho a} r^{2m+1}\,dr \qquad \text{(B12)}$$

Hence,

$$\int_0^{\rho a} r\, J_m^2(k_0 r)\, dr \simeq \frac{k_0^{2m}(\rho a)^{2m+2}}{(2^{2m}(m!)^2} \qquad \text{(B13)}$$

Equations (B4) and (B13) then give the small-$\rho$ scaling

$$s_m^{\text{cloak}} \simeq C_m(\epsilon_{\text{obj}} - 1)\frac{k_0^{2m}(\rho a)^{2m+2}}{(2^{2m}(m!)^2} \qquad \text{(B14)}$$

and

$$\partial_{\epsilon_{\text{obj}}} s_m^{\text{cloak}} \simeq C_m \frac{k_0^{2m}(\rho a)^{2m+2}}{(2^{2m}(m!)^2} \qquad \text{(B15)}$$

For the dominant monopole contribution $m = 0$, these relations reduce to

$$s_0^{\text{cloak}} \simeq C_0(\epsilon_{\text{obj}} - 1)\frac{(\rho a)^2}{2} \qquad \text{(B16)}$$

and

$$\partial_{\epsilon_{\text{obj}}} s_0^{\text{cloak}} \simeq C_0 \frac{(\rho a)^2}{2} \qquad \text{(B17)}$$

The corresponding classical scattering measure is proportional to the modal intensity sum

$$\sigma_{\text{sca}}^{\text{cloak}} \propto \sum_m w_m \mid s_m^{\text{cloak}} \mid^2 \qquad \text{(B18)}$$

where $w_0 = 1$ and $w_m = 2$ for $m > 0$. In contrast, the displacement-level QFI contribution in a detected modal subspace is proportional to the squared parameter derivative,

$$F_\mu^{\text{det,cloak}} \propto \sum_{m\in\text{det}} w_m \mid \partial_{\epsilon_{\text{obj}}} s_m^{\text{cloak}} \mid^2 \qquad \text{(B19)}$$

up to the detected noise normalization. Equations (B18) and (B19) are the relations used in Section 5 to distinguish classical scattering suppression from local quantum-state sensitivity.

The ideal singular-cloak limit corresponds to $\rho \to 0$, in which both $s_m^{\text{cloak}}$ and $\partial_{\epsilon_{\text{obj}}} s_m^{\text{cloak}}$ vanish algebraically. However, for every finite regularization parameter $\rho > 0$, the derivative in Eq. (B9) remains nonzero within the Born-limit benchmark. This is the sense in which a regularized cloak can suppress the scattering amplitude while still retaining a finite local derivative imprint of the hidden-object parameter.